\PassOptionsToPackage{unicode}{hyperref}
\PassOptionsToPackage{hyphens}{url}
\documentclass[12pt]{article}
\usepackage{amsmath}
\usepackage{graphicx,psfrag,epsf}
\usepackage{enumerate}
\usepackage[]{natbib}
\usepackage{textcomp}

\newcommand{\blind}{0}

\IfFileExists{bookmark.sty}{\usepackage{bookmark}}{\usepackage{hyperref}}
\IfFileExists{xurl.sty}{\usepackage{xurl}}{} 
\hypersetup{
  pdftitle={Tinkering Against Scaling},
  pdfkeywords={Scaling, Critical Studies, Computational Social
Sciences, LLM, Tinkering},
  hidelinks,
  pdfcreator={LaTeX via pandoc}}

\begin{document}

\def\spacingset#1{\renewcommand{\baselinestretch}%
{#1}\small\normalsize} \spacingset{1}


\if0\blind
{
  \title{\bf Tinkering Against Scaling}

  \author{
        Bolun Zhang \\
    Zhejiang University\\
     and \\     Yang Shen \\
    Zhejiang University\\
     and \\     Linzhuo Li \\
    Zhejiang University\\
     and \\     Yu Ji \\
    The University of Chicago\\
     and \\     Di Wu \\
    Zhejiang University\\
     and \\     Tongyu Wu \\
    Zhejiang University\\
     and \\     Lianghao Dai \\
    Zhejiang University\\
      }
  \maketitle
} \fi

\if1\blind
{
  \bigskip
  \bigskip
  \bigskip
  \begin{center}
    {\LARGE\bf Tinkering Against Scaling}
  \end{center}
  \medskip
} \fi

\bigskip
\begin{abstract}
The ascent of scaling in artificial intelligence research has
revolutionized the field over the past decade, yet it presents
significant challenges for academic researchers, particularly in
computational social science and critical algorithm studies. The
dominance of large language models, characterized by their extensive
parameters and costly training processes, creates a disparity where only
industry-affiliated researchers can access these resources. This
imbalance restricts academic researchers from fully understanding their
tools, leading to issues like reproducibility in computational social
science and a reliance on black-box metaphors in critical studies. To
address these challenges, we propose a ``tinkering'' approach that is
inspired by existing works. This method involves engaging with smaller
models or components that are manageable for ordinary researchers,
fostering hands-on interaction with algorithms. We argue that tinkering
is both a way of making and knowing for computational social science and
a way of knowing for critical studies, and fundamentally, it is a way of
caring that has broader implications for both fields.
\end{abstract}

\noindent%

\vfill

\newpage
\spacingset{1.9} 

\section{Introduction: the curse of
scaling}\label{introduction-the-curse-of-scaling}

``Scaling'' has emerged as the dominant paradigm in artificial
intelligence---more narrowly, in machine learning---over the past
decade. This trajectory began with Fei-Fei Li's bold initiative to build
the ImageNet dataset, was validated by the breakthrough success of
AlexNet, and ultimately reached its apex with the advent of large
language models. Scaling became the only way forward as it has been
naturalized as a law: \citet{kaplan_scaling_2020} observed from
experiments that neural network based models' performance can be
predicted using the number of model parameters, the size of the training
data, and the cost of model training. When all three factors grow
proportionally, the model's performance will also be better.
\citet{sutton_bitter_2019} crowned scaling in a Whiggish way: ``only
methods that leverage computation is the only thing that matters in the
long run, and specific human knowledge only helps in short-terms.'' Even
when prominent scholars like Ilya Sutskever reflect on it, it is not
such scaling law that hits the wall, but Moore's law comes to an end or
we already ran out of data.In this way, algorithmic techniques like
transformers quickly took the reign. ``Human centric'' computing is
dead, declared ``scaling''.

Blessing for some, scaling is a curse for others. AI/ML research used to
flourish on a symbiosis of academics and industry, based on a knowledge
infrastructure supported by commercial hardware and open-source
projects. This symbiosis began to break down during the hyper-scaling
age of large generative models: the model simply became too big for
academic researchers to train within academic institutions. As guessed,
GPT-4o might have 1.7 trillions parameters. This is 18000 times larger
than the AlexNetM that restarted the age of the neural network, and 5000
times larger than the original BERT-Large in which the transformer
architecture was first implemented in NLP. Training modern language
models is too expensive for a general research team to afford. Even
``small'' language models like the Phi series, would take at least 4
million US dollars to train. Besides financial capacity, training a
large model, or more precisely, leveraging computation power at scale
needs engineering expertise, such as accelerated communications and
virtual network among large amounts of GPUs, is usually absent in the
academics, similar to the infrastructure itself
\citep{kudiabor_ais_2024}.

If such an imbalance has bothered computer science researchers in
general, it creates dual challenges for social sciences: one for
computational social science researchers who want to tame the machine
learning algorithms to explore the social world, one for more critical
group of researchers in digital sociology, critical algorithm studies,
and other similar fields, who seek to interrogate and understand the
epistemic limits and methodological constraints when engaging with
opaque, large-scale algorithmic systems. What's more, both groups must
the confront structural inequalities and sociotechnical risks embedded
in algorithmic systems deployed by major tech firms and public
institutions while their capacities to do so are unevenly distributed.

For computational social science researchers, their capacities were
greatly augmented and limited by scaled up models at the same time. It
is undeniable that large language models are more capable than existing
tools in computational social sciences, which themselves were imported
from computer sciences and industry as well. Meanwhile, they also
transform researchers' roles. Using the new tools, researchers can
hardly control every step of their research processes. They used to be
full-stack developers: theoretically, they can understand and control
every single step of the modeling process, from data collections, to
training and evaluating the models. In front of the large language
models, especially those close-sourced state-of-art models, they know
little about the training details, nor to say how to ``control'' them.
At best, they can be reviewers of these models via designing benchmarks
specifically for social science tasks. When they utilize these models,
they can hardly ensure the reproducibility of their results, as
commercial entities like OpenAI continuously retire their older models
\citep[For a recent example, see][]{bisbee_synthetic_2024}. If the
imbalance of logistic power and expertise poses a pressing issue for
computer science researchers, it bothers computational social scientists
even more.

In critical algorithm studies and investigations of algorithmic
systems/societies in general, hyper-scaling also introduces a problem of
knowing. Due to a general lack of expertise in the social sciences,
critical researchers increasingly lack a broad understanding of
algorithmic systems in the wild and have been persistently locked into
the metaphor of the ``algorithm as a black box''
\citep{seaver_knowing_2019}. As a consequence, many works tend to offer
purely external criticisms of algorithmic technologies, attributing an
all-encompassing and overwhelming causal power to algorithms in their
narratives \citep{seaver_knowing_2019, christin_ethnographer_2020}. Many
works have treated algorithms merely as a contextual backdrop and failed
to provide specific empirical details for further interpretation or
explanation.

In both sub-fields, hyper-scaling also creates a new hierarchical
principle based on researchers' proximity to big tech companies, where
data, computing power, and expertise are concentrated. Those who are
closer to these companies and have access to these resources become the
privileged class of researchers. When computational social science was
first conceptualized more than a decade ago,
\citet{lazer_computational_2009} warned that if computational social
science did not strive to be an open science, then researchers
affiliated with big internet companies and who have access to non-public
data would become a privileged group whose work could not be reproduced
or critiqued. Now, this possibility has become even more apparent and
pressing under hyper-scaling, given that not only data but also
computing power and expertise disproportionately reside in the industry.

Ironically, a similar trend appears in critical research. Since the
``black box'' metaphor reconfigures the issue of knowledge into a
problem of ``access''
\citep{burrell_how_2016, seaver_knowing_2019, christin_ethnographer_2020},
companies then try to lure these critical scholars with it. Big tech
companies in the U.S. and China are trying to recruit researchers in
this field into their networks for various causes. Sometimes they do
express concerns related to the social impact of algorithms, but more
unmistakably, there are also elements of public relations at play.
\citet{pasquale_two_2016} used to distinguish researchers' stances
toward platform companies---which are also the organizers of many of the
most important algorithmic systems---into two groups: one optimistic and
the other more critical. Although such official ``access'' projects did
not produce the same dichotomy, companies certainly favor certain types
of these projects. Whether a researcher can maintain their independence
under such corporate affiliations depends on the specific arrangements
of the project framework. However, we should be concerned given that
individual researchers usually have little bargaining power.

These parallel dynamics---unequal access to data, expertise, and
infrastructure; the co-optation of critique; and the epistemic limits
imposed by the black-box metaphor---point to a need for alternative
modes of engagement with algorithmic systems. Rather than striving to
scale up or gain privileged access to industrial infrastructures, we
suggest that researchers might instead look sideways: toward smaller,
slower, hands-on, and more materially grounded practices. In this
spirit, we propose a methodological agenda centered around tinkering
---or put more bluntly, the act of playing with and modifying smaller
models or components that ordinary researchers can handle, both
technically and materially. This calls for a new positionality toward
algorithmic tools in the social sciences. We argue that tinkering lies
at the crux of bridging computational social science and critical
studies of algorithmic societies.

To articulate this framework, we first discuss the theoretical
inspirations for tinkering, drawing from philosophy, history, and
science and technology studies (STS). We then provide two sets of
examples to illustrate how tinkering can function both as a way of
knowing in critical investigations of algorithms and as a mode of
enacting/making in computational social science. We conclude with a
discussion of how tinkering can also be understood as a form of caring:
a normative approach for repositioning the humanities and social
sciences in the face of the challenges posed by large generative models.

\section{Theoretical inspirations}\label{theoretical-inspirations}

\subsection{Releasement and
reflection}\label{releasement-and-reflection}

Our major source of inspiration comes from Heidegger's discussion on
technologies\citep{heidegger_question_2013, heidegger_traditional_1998}.
Here, our purpose is not to talk about Heidegger's writing for its own
sake---this is a task for philosophers. Instead, we aim to appropriate
Heidegger's insights into technology in the same way he appropriated
Ancient Greek philosophy: namely, to adjust our questioning of
technology to the scale of the questions of human existence, while
simultaneously remaining grounded in our current experience of
technologies.

For Heidegger, ``modern technology'', or ``what is technological'', is
not defined by its usage or technical features---whether a machine is
powered by combustion engines or electric motors---but by its unique way
of world-making. This world-making, which Heidegger calls ``bringing
forth,'' refers to the process by which an existence, whether a thing or
human, realizes its potential and becomes ``unconcealed.'' In this act
of revealing, the potential inherent in an existence is made apparent,
bringing that existence into the open.

Unlike other forms of ``bringing forth,'' technology enacts a mode of
``challenging forth''(Herausfordern). In this process, the objectness of
things under the control and manipulation of technology is
predetermined, and alternative possibilities are systematically closed
off. Heidegger illustrates this with the example of the hydroelectric
dam: when the dam is built on the Rhine, the river is no longer the one
in Hölderlin's hymn, but is reduced to ``a water power supplier, derives
from out of the essence of the power
station.''\citep[16]{heidegger_question_2013} In this sense, the Rhine,
stripped of all its other potential modes of existence, is reduced into
a ``standing-reserve'' for electricity in a power grid. In short,
technology, ``drives out every other possibility of
revealing.''\citep[27]{heidegger_question_2013}

This reduction of existences to standing-reserves is not confined to the
natural world. Modern technologies challenge everything, including
humans, to become standing-reserves. This is what Heidegger refers to as
``enframing'' (Gestell). In his critique of early natural language
processing techniques, this point becomes even more apparent. For
Heidegger, language is not merely a tool; it is a fundamental part of
human essence. We are not masters of language, as language is not
something we control; rather, ``language is the house of being.''
Following Wilhelm von Humboldt, Heidegger writes that ``language is the
between-world between the human being's spirit and objects,'' serving as
an expression ``of this between, between subject and
object.''\citep[139]{heidegger_traditional_1998}

This view contrasts significantly with early information theory and the
early NLP techniques, as represented by Norbert Wiener. Wiener defined
learning ``in its essence {[}as{]} a form of feedback, in which the
pattern of behavior is modified by past experience.''
\citep[141]{heidegger_traditional_1998}. In this conception, language is
reduced to ``a mere report of signal transmissions,''
\citep[140]{heidegger_traditional_1998} ---information that can be
processed and acted upon by machines. Heidegger calls this reduced form
of language ``technical language.'' It is not difficult to see that this
early version of NLP shares affinities with the current paradigm of
large language models.

Under the hyper-scaling of large language models, the distinction
between humans and machines becomes increasingly blurred. To use
Heidegger's analogy, we humans are becoming standing reserves of
technical languages for these generative models. The rise of large
language models marks a major shift in the sociotechnical imagination of
data. Previously, data was likened to oil---a resource to be extracted.
Unlike oil, data does not deplete through use; it can be recycled
endlessly in model training. However, under hyper-scaling, the vast
amount of data needed for training these models is anticipated to be
exhaustible \citep{sutskever_sequence_2024}. Crucially, in this
hyper-scaling paradigm, language---once a medium for human creation and
communication---has now become a standing reserve for ``machine
capacities.''

Confronted with this challenge concerning the meaning of being,
Heidegger does not advocate a total rejection of modern technology;
rather, he calls for a shift in our positionality---a change in how we
relate to technology, or, in his term, other ways of ``bringing forth.''
In his later work, Heidegger introduced the concept of ``letting be'' or
``releasement'' (Gelassenheit), which offers a more reflective and
balanced way of relating to technology. ``We let technical devices enter
our daily life,'' Heidegger suggests, ``and at the same time leave them
outside, letting them alone as things which are not absolute but remain
dependent upon something higher.''

This releasement is not passive acceptance; it requires an active,
reflective stance. For Heidegger, modern technologies and the
calculative thinking behind them have driven humanity into a
``homeless'' state. To retrieve the essence of humanity, we must bring
reflection back into our relationship with technology. True releasement
requires that we engage with technology in a way that is not simply
functional but also reflective---questioning its role and purpose in our
lives. Heidegger traces this back to the Greek origin of techne, which
means ``to know one's way in something,'' particularly in the act of
creation or production (poiesis). This knowledge is not mere technical
expertise, but an engagement with the essence of the thing being
created. The freedom in technological dominance requires such a double
revealing. It aims at bringing the thing into the unconcealment, but
also ``let the veil'', which causes the conceals in the first place,
``appears at what veils'' \citep{heidegger_question_2013}.

This ``active releasement'' brings us to the concept of tinkering. As we
navigate the overwhelming scale of technological advancement, we must
resist the assumption that accumulative scaling is the only path
forward. Tinkering offers a way to engage with technology that is
grounded in reflection, local practice, and adaptability through
interacting with technology on a more intimate and contextual
level---rethinking the tools we use, and how we use them. In this way,
tinkering becomes a form of resistance to the unchecked myth of
hyper-scaling, and an attempt to restore balance between human needs and
technological capacities. In Heidegger's term, it is a
``double-revealing.''

\subsection{Tinkering as a
double-revealing}\label{tinkering-as-a-double-revealing}

The key to Heidegger's reflection on technology is not so much about
acquiring new knowledge regarding machines and tools themselves, as it
is about re-establishing a meaningful relationship with the essence of
technology. In the face of technological dominance---specifically its
``challenging forth''---the challenge that both Heidegger and we
specifically take up is to restore a space of freedom for more open
human participation. This freedom is not achieved by rejecting
technology outright, but by re-formulating our positionality in relation
to it. This requires us to recognize that we have already been
integrated into the technological apparatus, subject to the same
calculative forces that govern the functioning of machines and systems.
Yet at the same time, by activating technology's capacity for
revealing---at the edge where its very enframing occurs, where Heidegger
sees the ``saving power'' residing in the ``danger''---we claim the
irreducible part of reflective freedom in relation to the condition of
scaling technologies.

This is where tinkering comes in a potential form of ``double
revealing.'' We suggest that tinkering offers a way of engaging with
technology that embodies this dual movement: revealing the technology
itself while also revealing its limitations and alternatives. In this
process, we do not merely submit to the technological forces shaping our
world; instead, we actively engage with these technologies, letting them
incrementally work towards direction and form an acceptable arrangement
under local constraints \citep{jansky_digitized_2024}. In other words,
where freedom does not reside aside from technology but within
technology's entanglement with the worlding of our world. The freedom
granted by tinkering is to re-define how technology reveals itself to us
and how we reveal ourselves through it. Tinkering as an iterative
practice allows us to reflect on both the technology and our
relationship to it. It is not about finding a perfect solution; rather,
but through making small adjustments, we re-create a co-habitable world
in which we all find ourselves in one way or another.

``Tinkering'' in STS is a concept developed through STS studies of care
practices, and it can be traced to many classical works in anthropology
\citep{levi-strauss_savage_1966, mol_logic_2008, mol_care_2010, ingold_making_2013}.
In diabetes care, patients, nurses, and doctors do not necessarily
choose the best blood sugar monitor, but rather the one that fits their
life rhythms. The same applies to their choices around insulin
injections \citep{mol_logic_2008}. \citet{jansky_digitized_2024} found
that individuals with Type 1 Diabetes (T1D) engaged with open-source
automated insulin delivery systems within a community in order to
personalize and improve their self-care, compensating for relatively
outdated medical technology in Germany. Through this engagement,
participants not only acquired knowledge about the technical details of
the system and their own bodies, but also contributed local improvements
to the broader community. Therefore, tinkering involves working with
technologies while refusing to accept them uncritically. It is oriented
toward specific goals and generates cultural meanings in the process.

This concept is also mobilized in recent histories of technology
concerning vernacular capitalism and local innovation, challenging
conventional understandings of industrial development. Similarly,
tinkering here refers to copying, experimenting with, and improving
existing technologies rather than inventing from scratch. Through the
stories of figures like Hugo Gernsback and Chen Diexian---both writers
and avid tinkerers---historians propose tinkering as an alternative to
mass technologies and the dominant development paradigms behind them
\citep{wythoff_introduction_2016, lean_vernacular_2020}.

Here, we suggest that researchers themselves should become tinkerers of
models to resist the dominance of hyper-scaling. In practice, tinkering
encompasses a set of existing modeling practices in the social sciences.
As we previously mentioned, what we advocate is a more conscious and
systematic turn---a shift in positionality. We distinguish tinkering
across three layers:

For one thing, it emphasizes the value of small models, which allow
researchers to reflect on every step of the modeling process.

Next, it involves engaging with larger pre-trained models, especially
open-source ones. This layer ranges from basic prompt
engineering\footnote{In practice, this can be highly technical and
  complex. Prompt engineering utilizes the context window of generative
  models and revives many elements of rule-based approaches.} to probing
the inner workings of large models and fine-tuning them for specific
tasks.

Finally, tinkering also refers to conceptualizing and designing models
tailored to the needs of the social sciences. This includes close
reading of computer science literature and appropriating foundational
components of large models---such as the transformer
architecture---while endowing them with new cultural significance.

Tinkering, thus, assumes a different set of relations with models.

First, echoing Heidegger's idea that unconcealment should not be forced,
we see ourselves as negotiating with the model rather than manipulating
or controlling it to produce outcomes we crave for. Relatedly,
data---and the social lives embedded within it---should not be treated
as standing reserves to be mined or extracted.

Second, modeling as a form of double revealing suggests that we gain not
only new knowledge from the model's outputs but also insights from the
modeling process itself. Modeling is a form of knowing: through it, we
come to understand not only our methods more deeply, but also the data.
This view calls for a reconsideration of the rigid divide between
methods and theory, which are, in fact, always already entangled.

Lastly, for critical researchers, tinkering---which must begin with
tangible materials---offers a new epistemic starting point and prompts a
closer engagement with the ``metal,'' not just the ``mental.'' Although
the modeling process may differ significantly from the algorithmic
systems operated by large corporations, it can function as a toy
model---an ideal type that supports both comparison and critique.

However, highlighting tinkering also reveals some major differences
between our position and that of Heidegger, which are worth addressing
here. Most importantly, unlike Heidegger, we do not place words or
speech in a position of ultimate sublimity. For Heidegger,
mediation---especially that which is non-instrumental and not oriented
toward a goal---is more closely aligned with the essence of the human.
Language, for him, exemplifies this mediation in the disclosing of
Being. In contrast, we argue that making---when approached not as a
means to an end but as an exploratory, open-ended engagement---can
itself be a form of reflective mediation. Tinkering, as we propose it,
does not reject Heidegger's concern with non-instrumental thinking, but
relocates it in material practice. Like language, tools and models
mediate our relation to the world, and engaging with them can produce
insight, hesitation, and care. A similar divergence emerges in relation
to Arendt, who, while sharing Heidegger's critique of calculative
thinking , locates the political in speech and action rather than in
making. In her understanding of the human condition, she explicitly
distinguishes labor and work from action, assigning the latter a
uniquely political function \citep{arendt_human_2019}. In this view,
politics has little to do with making or technology. In contrast, we
contend that politics is not solely discursive or speech-action-based;
it is also materially enacted and partially constituted by the ordering
of things \citep{law_globalisation_2008}. Making---when approached not
as a means to an end but as an exploratory, open-ended engagement---can
itself be a form of reflective mediation. In other words, rather than
treating words and tools as opposites, we may reconsider the boundaries
between the two and understand both as mediations between humans and the
world. And rather than offering a critique of the historical condition
of our time in the style of Heidegger or Arendt., we instead propose
tinkering as a modest practice for local world-making --- a way to
subtly reconfigure the material-conceptual order imposed by
hyper-scaling.

In short, tinkering represents a non-antagonistic positionality toward
algorithmic technologies such as large language models, while
simultaneously asserting that scaling is not the only path forward. In
what follows, we present two sets of examples from existing
research---one rooted in critical scholarship, the other in
computational social science (CSS)---to demonstrate the feasibility and
potential of this framework.

\section{Tinkering in computational social
sciences}\label{tinkering-in-computational-social-sciences}

Computational social sciences usually import algorithms from their
colleagues in computer sciences, and take a little consideration about
the difference between the industry and the academics. After all, the
later group has a much tighter relationship with the industry.

Algorithms under the modern machine learning framework usually have a
certain intentionality built-in
\citep{breiman_statistical_2001, wallach_computational_2018}: they are
designed to make predictive associations. In industry, these predictive
associations were crucial for making profits. These algorithms are not
designed to explore our social world.

Because of this, many issues that won't be a problem for industrial
usages might be troublesome for social scientists. For example, when
using topic models, a widely used unsupervised machine learning
algorithm, a researcher would find that there are many different local
solutions that have similar topics, while these topics of different
local solutions might yield contradictory conclusions when applied in
downstream statistical tests. This won't be an issue in the industry as
long as the model outperforms its predecessors in certain KPIs on large
scale \citep{zhang_sts_2025}.

Tinkering, in this context, is a mode of translation, to re-situate the
methods from other fields to better align with the aims of social
science and the exploration of the social world.

\subsection{Tinkering small models}\label{tinkering-small-models}

The first type of tinkering we want to discuss involves making better
use of smaller models. ``Small'' here is a relative term---compared to
hyper-scaled models, these models don't demand massive infrastructure.
Researchers can work with them using modest resources, such as a small
university cluster, an affordable cloud computing setup, or even a
personal computer.

For example, to address the limitations of topic models discussed
earlier, \citet{zhang_can_2024} proposed a method called the Similar
Topics Identification Procedure (STIP) to enhance their application in
the social sciences. They argue that after identifying a reference
model---one deemed appropriate for answering the research
question---researchers should also explore the broader model space to
identify alternative models that are semantically similar to the
reference. To do so, a variety of research designs can be employed. In
their demonstration, \citet{zhang_can_2024} used the Average Jaccard
Index on distinguishing terms to detect sets of semantically similar
solutions. Once identified, researchers can examine whether these
similar models produce conflicting outcomes in downstream statistical
analyses. If discrepancies arise, they can then investigate potential
causes and refine the validation process for their models.

\begin{figure}
\includegraphics[width=0.9\linewidth]{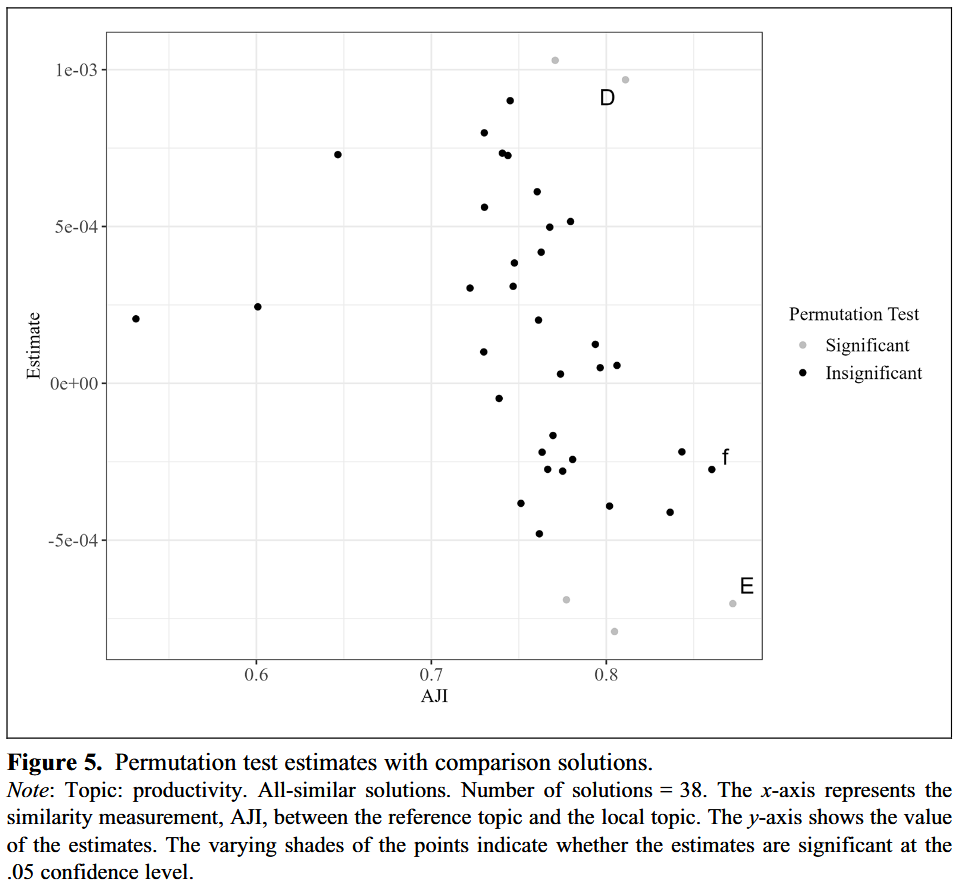} \caption{Applying STIP on a topic from Fligstein et al. 2016. Y axis presents the estimates and X axis presents the AJI, a similarity measurement. We found that even among topics highly resemble the reference model, there are opposite results.}\label{fig:fig1}
\end{figure}

STIP aligns with a growing number of methods in computational social
science that adopt a model-verse approach
\citep{chuang_topiccheck_2015, munoz_we_2018, knox_testing_2022}. This
perspective resonates with the agnostic view of modeling, which holds
that each model offers a partial and angled lens on the social world
\citep{grimmer_machine_2021}. By employing multiple models rather than
relying on a single one, researchers can begin to assess model-level
uncertainty more systematically.

Perhaps more important for our purposes is how the model-verse approach
highlights a key difference in inferential priorities between academia
and industry that can be utilized by tinkering. The key issue of
inference in the industry, using the words of Michael Jordan, one of the
LDA topic model's designers, is ``can you generate a certain level of
inferential accuracy within a certain time budget even as the data grow
in size?''\citep{jordan_statistics_2013} For example, when topic models
are used in online advertising to support predictive associations or
collaborative filtering, inference must happen almost
instantaneously---an ad must be selected and displayed as a webpage
loads, typically within seconds. Industry, therefore, cannot afford to
explore the broader model space in business operations. In contrast,
social scientists operate under fewer time constraints, but face higher
stakes regarding the validity and interpretability of their results.
This allows them to take advantage of the model-verse approach by
comparing multiple models rather than relying on a single one.

Given that many algorithms have yet to be fully optimized for social
science applications, there remains ample space for researchers to
tinker with small models. Taking topic modeling again as an example,
social scientists can revisit core concepts like overfitting and
smoothing from the perspective of interpretability. They might also
apply additional techniques such as coefficient pruning to strike a
better balance between human readability and measurement accuracy.

\subsection{Probing into larger
models}\label{probing-into-larger-models}

The second form of tinkering is to probe into large models, especially
recent large generative models. We are witnessing a quick adoption of
large language models in social sciences, especially in the task of
coding \citep{ziems_can_2024, bail_can_2024a}. One main issue of
generative models like large language models is hallucination, i.e.,
models generate information that appears plausible but is factually
incorrect or nonexistent. We do not know in general whether this is
because the model over-fits biased training data or because there is no
related information in the training data. A common approach to it is to
assess the model's performance in specific tasks under the supervised
machine learning framework, and compare the model's result to human
coding. Recent work also suggests that via combining random sampling and
expert coding, researchers can further adjust the results through
weighting \citep{egami_using_2023}.

\citet{li_old_2025} devices a different strategy. He argues that instead
of assessing the results afterwards, we should also probe into the
model. Li mobilizes the metaphor of LLM as an average interviewee on the
internet, and adopts techniques in conducting survey.

As \citet{li_old_2025} indicated, while large language models offer
powerful capabilities for text analysis at scale, their reliability
remains a pressing concern. Traditional approaches like expert
validation \citep{egami_using_2023} become impractical at scale, yet
blindly trusting model outputs risks compromising research integrity
\citep{zhang2023safetybench, tornberg2024best, palmer2024using, ollion2024dangers}.
This is where tinkering through probing becomes essential - not just to
validate results, but to understand the models' behavioral patterns and
limitations.

One promising direction is to draw insights from established
methodologies in social sciences. For instance, survey methodology's
rich tradition of addressing respondent behavior offers valuable
frameworks for probing LLMs. Just as survey researchers developed
techniques to detect satisficing (\citet{krosnick_response_1991}) -
where respondents provide superficially satisfactory but not fully
considered answers - similar approaches can be adapted to probe LLM
behavior. This allows researchers to move beyond treating models as
black boxes and instead engage with them through experimentation and
observation.

\citet{li_old_2025}'s experiments focused on classifying scientific
papers of faculty1000 into three major contribution types: ``Interesting
Hypothesis,'' ``Technical Advance,'' and ``New Finding'' based on the
text of paper abstracts. Three different sizes of Llama 3.1 Instruct
models (with 8B, 70B, and 405B parameters) are employed under analysis.
The task is set in such a way that LLM's only need to predict one option
token after presented with a set of choices along with paper abstracts.
This classification task is a particularly suitable case for probing
model behavior as it exemplifies a common use case for AI to make
categorical choices. Li implemented three classic strategies of
survey-inspired interventions: option randomization (varying the order
of response choices), position randomization (altering the prompt
structure in context), and reverse validation (using inversely coded
questions), and examined the change of probabilistic distributions of
LLM prediction of answer token. The results revealed significant
instability in model responses. The flip rates (changes in model
predictions) varied notably across model sizes and paper categories (see
Figure 1), with smaller models showing higher sensitivity - flip rates
reaching 25.4\% under certain interventions. More intriguingly, even the
largest 405B parameter model exhibited non-negligible response changes.
The instability was particularly pronounced for less frequent categories
(see Figure 2) - papers classified as ``Interesting Hypothesis'' showed
higher flip rates across all interventions and model sizes compared to
the more common ``New Finding'' category.

\begin{figure}
\centering
\includegraphics{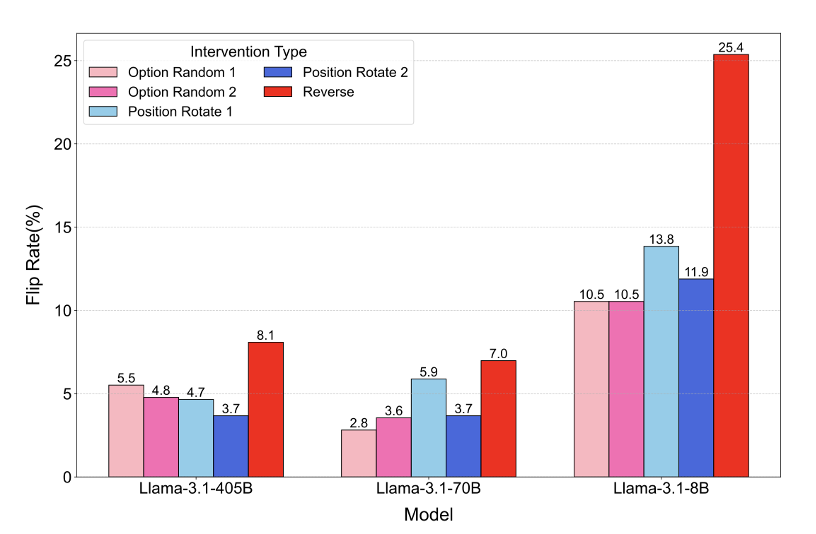}
\caption{Impact of survey-inspired interventions on LLM annotation
reliability. Adapted from Figure 3 in \citet{li_old_2025}.}
\end{figure}

Figure 2. above presents the flip rates of LLM answers under three
survey-methodology interventions across different model sizes. Higher
flip rates indicate greater annotation instability under different
prompting conditions.

\begin{figure}
\centering
\includegraphics{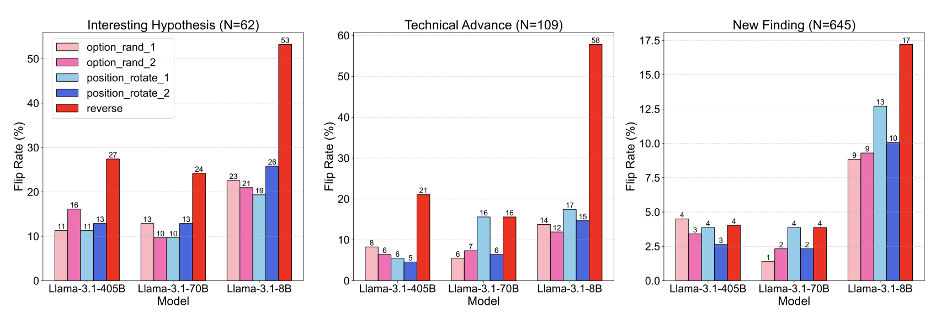}
\caption{Category-specific impacts of interventions on LLM annotation
reliability. Adapted from Figure 4 in \citet{li_old_2025}.}
\end{figure}

Figure 3's results show varying levels of annotation stability across
different paper categories (Interesting Hypothesis, Technical Advance,
and New Finding).

These findings highlight a fundamental perspective that social
scientists should adopt when working with LLMs: these models are
inherently probabilistic rather than deterministic machines. Instead of
treating model outputs as fixed, definitive answers, researchers should
consider them as probability distributions over possible responses, each
carrying different levels of confidence and uncertainty. This
probabilistic nature becomes evident through the probing experiments in
Li's paper \citep{li_old_2025}, which yield several suggestions for
social scientists. First, model size alone doesn't guarantee reliability
- even larger models can exhibit inconsistent behavior under different
prompting conditions. Second, content type matters significantly - the
same model may be more reliable for common categories but struggle with
rare ones. Most importantly, systematic probing can reveal potential
weaknesses in the model's annotation capabilities that might otherwise
go undetected. By adopting this probabilistic perspective and employing
systematic probing techniques, researchers can better assess and account
for the inherent uncertainties in LLM-based analyses.

Through such methodical probing and tinkering, social scientists can
play a crucial role in developing better practices for AI use, both in
academic research and public applications. Li's work \citep{li_old_2025}
also exemplifies how traditional social science knowledge can be
repurposed to study and improve new technological tool.

Again, probing into the model like Li does would also offer us a gauge
of uncertainty to some extent. It also highlights a general attitude to
generative models: the more we emphasize on the probabilistic aspect
rather than the generative aspect of the model, the less hallucination
we are going to have.

Extending this line, social scientists can and should play a bigger role
in model interpretability in machine learning. Every model, including
LLMs, contains hidden assumptions about how they represent the social
and cultural world. These assumptions are embedded in their
architectures, training processes, and the data they learn from,
inevitably shaping their predictions and behaviors. To unveil the
consequences (good or bad) requires making model behaviors more
interpretable. Currently, model interpretability studies are mainly
restricted within a small community in the deep learning field,
exemplified as ``mechanistic interpretability''(MI) studies. For
instance, Anthropic's research teams have pioneered work on mechanistic
interpretability of transformer models, attempting to understand the
internal representations and induction making processes of simple
transformer models \citep{elhage2021mathematical}. Related technical
approaches such as attribution methods
\citep[@rauker\_transparent\_2023,@wang\_interpretability\_2022]{sundararajan_axiomatic_2017}
are being developed. However, despite their progress, these studies
primarily approach interpretability from a technical perspective,
focusing on computational mechanisms rather than their social and
cultural implications, and they haven't demonstrated their scalability
with more sophisticated analysis. Recent advances in reasoning AI such
as Deepseeks' R1-zero, hinted both at the possibility of increasing
difficulty in interpretability, as AI can now be trained to emerge chain
of thoughts in the form of mixed or even nonhuman languages
\citep{deepseek-ai_deepseekr1_2025}, and new opportunities for
interpretability, as now AI can show explicit ``chain of thoughts''
before answering. As LLMs become more powerful, we believe social
scientists are obligated to integrate MI methods with social and
cultural analysis.

\subsection{New Modeling
Methodologies}\label{new-modeling-methodologies}

Lastly, we suggest that tinkering here also includes appropriating
components of larger models, giving them new cultural significance and
devising new models specifically for social sciences and humanities. In
other words, we will bring human knowledge back-in, instead of relying
on ``larger and larger steam machines'' to harness more computing power.

According to \citet{ji_metasemanticmetapragmatic_2025}, current
mainstream approaches to designing and evaluating multimodal
communicative AI fail to capture the complexity of human multimodal
communication. This issue cannot be resolved simply through increased
parameterization, larger datasets, or post-training methods that neglect
the rich and diverse interpretive dimensions of human communication.
Advancing multimodal AI research requires a more rigorous
conceptualization of modality---one that moves beyond its material
representations (text, sound, visual) and interrogates its role within
communicative interactions.

Rather than relying solely on scale and conflating mappings between
perceptual-sensory, sociocultural, and reasoning domains,
\citet{ji_metasemanticmetapragmatic_2025} proposes a modeling
methodology for multimodal communicative alignment that are sensitive to
both symmetrical and asymmetrical shifts across domains and adapting to
contexts and situations. Based on this framework, typical multimodal
communicative AI models primarily optimize for representational
cross-modality coherence rather than communicative fluidity, leading to
ineffectiveness or failures in navigating context-sensitive, contingent,
or pragmatic shifts across modalities. Building on the foundational work
of Charles Sanders Peirce, as well as theories in contemporary
philosophy of language, sociolinguistics, and semiotic communicative
theories, this framework identifies three key modalities: iconic
(sensory and perceptual), indexical (relational, situational,
contextual), and rule-like (presupposing or entailing reasoning-based
inputs or outputs). Implementing a social-science-driven, relational and
context-sensitive model architecture requires explicitly encoding how
communicative relations and context is rendered salient or navigated
towards specific contextual direction in multimodal navigation---moving
beyond simple semantic coherence and material representations that
dominate current multimodal AI approaches.

Such models would incorporate components that track and adjust to
relational, contextual, and situational shifts across interactions,
ensuring that communicative cues for these transitions are neither
flattened into token-based predictions nor reduced to static
representational mappings between modalities in the fine-tuning or
post-training stages. To achieve this, training methods should
distinguish between context-aligned and non-context-aligned multimodal
interactions using contrastive loss objectives that penalize misaligned
contextualization attempts.

\begin{figure}
\centering
\includegraphics{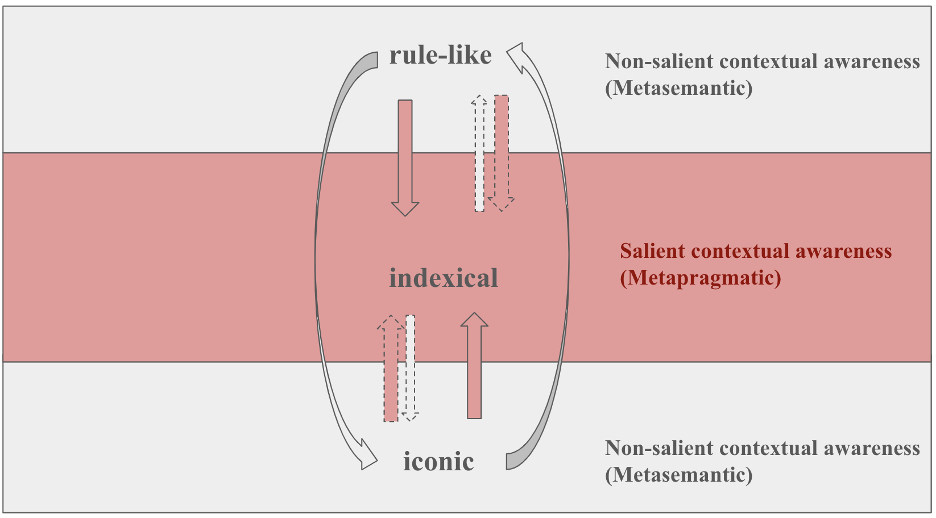}
\caption{Multimodaility in communication}
\end{figure}

Figure 4. above adapted from Figure 1 in
\citet{ji_metasemanticmetapragmatic_2025} Non-scalable contextualization
directionality: In the same or across socio-ecologically relatable
communicative scenes, once indexical encontextualization (light red
arrows) is performed, decontextualization (dotted gray arrows) becomes
increasingly difficult unless a proper recontextualization (dotted light
red arrows) is expected to be plausible. In contrast, transitions
between rule-like and iconic modes (curved gray arrows) do not exhibit
asymmetrical directionality, as both modes indicate similarly
non-salient contextual awareness.

This perspective underscores the need to establish new evaluative
benchmarks informed by social sciences, rather than relying solely on
narrowly defined multimodal benchmarks that merely scale up from
existing, self-referential metrics within computer science. Current
benchmarks for multimodal AI primarily assess performance through
specifically defined task-based success metrics---such as language-image
alignment or instruction-following accuracy---without accounting for the
full breadth and depth of human communicative competence. As a result,
multimodal alignment is often reduced to denotational-representational
mapping (as in language-image tasks), spoken language is flattened into
standardized written text, and versatile communicative functions are
constrained to goal-directed, service-like exchanges. We argue that
multimodal models should be benchmarked not only on their ability to
represent cross-modal content but also on their capacity for multimodal
contextual navigation and adaptation, drawing from much broader and more
realistic communicative benchmarks informed by contemporary social
sciences.

To enable this shift, \citet{ji_metasemanticmetapragmatic_2025} proposes
potential evaluative benchmarks for multimodal communicative models
based on their ability to:

\begin{enumerate}
\def\labelenumi{\arabic{enumi}.}
\item
  Characterize contextualization capacities as encontextualization,
  decontextualization, and recontextualization of multimodal inputs and
  outputs, in order to adapt to both semantic and pragmatic constraints
  (in linguistic modalities) as well as representational and
  non-representational factors (in non-linguistic modalities).
\item
  Based on (1), sustain, navigate, and shift between contextually
  appropriate communicative references by leveraging both flexibility
  and directionality in contextualization across diverse situational,
  social, cultural, and ecological contexts.
\end{enumerate}

Prioritizing the three-fold contextualization capacities and contextual
flexibility and directionality shifts the focus away from brute-force
scaling or post-tuning after large-scale pre-training for multimodal
communicative modeling, emphasizing instead an iterative,
interaction-sensitive, and interpretable approach.
\citet{ji_metasemanticmetapragmatic_2025} highlights that this
reorientation has profound implications for broader AI alignment,
particularly in three key areas. First, in intentionality modeling, it
ensures that AI systems can recognize and adapt to the relational and
pragmatic intent behind multimodal communication. Second, in cultural
and sociolinguistic adaptation, it enables models to move beyond
decontextualized datasets and develop the capacity for dynamic,
real-time adaptation. Finally, in ethical transparency, it fosters
systems that make communicative processes explicit, rather than relying
on opaque approximations of human relational and contextualization
capacities.

This modeling approach challenges the assumption that increasing data
and model size inherently leads to better AI. For AI systems designed to
navigate human communication, the goal should not be to surpass humans
in abstraction or efficiency but to develop models capable of tinkering,
adjusting, and engaging in communication as human interlocutors do,
whether contextually, adaptively, or interactively.

\section{Tinkering in critical studies of
algorithms}\label{tinkering-in-critical-studies-of-algorithms}

Equally important for our purpose, tinkering can serve for critical
research on algorithmic systems. For years, such research has often
revolved around the metaphor of the algorithm as a black box---something
to be opened in order to reveal hidden biases and structural
inequalities. But this framing, while politically urgent, encounters two
key limitations.

On the one hand, it underestimates the technical difficulties of
``opening'' predictive models. In modern machine learning, the goal is
not to model the true data generation process, but to maximize
predictive accuracy on new inputs. As a result, both the underlying data
and the algorithm often remain opaque. The internal parameters of a
model may ahve little to no interpretable relationships with real world
phenomena. This paradigm not only legitimizes the use of highly complex
models like deep neural networks, but also fragments the very notion of
interpretability, making it context-dependent and contested

On the other hand, this view also underestimates the organizational
processes behind the enacting of an algorithmic
system\citep{passi_problem_2019, seaver_knowing_2019}. As
\citet{seaver_knowing_2019} points out, algorithms are not developed in
a vacuum; they are continuously modified within organizational contexts.
Given that organizational goals must be translated into measurable
targets and encoded as variables for prediction, the same predictive
technologies are also held accountable to organizational key performance
indicators (KPIs). And this can differ significantly depending on
institutional settings and national contexts
\citep{christin_counting_2018}.

As we previously argued, under the hegemony of scaling, this black-box
understanding of algorithms becomes increasingly problematic. Algorithms
are often attributed with ultimate causal power, yet empirical studies
rarely specify or demonstrate how this causality unfolds in practice
\citep{rieder_political_2022}.

One way to move beyond the black-box metaphor is through tinkering.
First, tinkering equips researchers with a vocabulary to engage with the
specificities of empirical data. Second, it offers researchers new
perspectives for theorization.Perhaps most importantly, it creates
opportunities for generating cultural insights that go beyond critiques
and toward conceptual innovation. . Here we select a group of works that
we think tinkering has already played an important role, to show the
feasibility of this approach for critical research on algorithms.

\subsection{Retrieving specificities}\label{retrieving-specificities}

On the most apparent level, tinkering can help researchers to retrieve
the specificities in describing the functioning of algorithmic systems.

This has been approved as valuable already by Adrian Mackenzie's work on
the framework of machine learning. His book, \emph{Machine Learner}, is
of an interesting kind. The title of this book makes a pun: machine
learner not only refers to machine learning algorithms, but Mackenzie
himself. Different from observing from outside like many anthropologists
would do, nor reading computer science papers as foreign documents,
Mackenzie played with the data and codes with help of infrastructures on
various scales to ``occupying operational subject positions'', something
we would classify as tinkering in this paper. He sees machine learning
as a ``operational formation'', which is parallel to what Foucault
called ``discursive formation'', and dissects it diagrammatically to six
major components: vectorization, optimization, probabillzation, pattern
recognition, regularization and propagation. What makes this work unique
is that through tinkering, Mackenzie created a proximity between codes
and writing. This allows him to produce richer details without
``preemptively ascribing potency to mathematics or algorithms.''
\citep[9]{mackenzie_machine_2017}

A similar approach is taken by Bernhard Rieder, a media scholar and
tinkerer. One of his major concerns is that ``the critical analysis of
concrete technical objects, procedures, and practices is exceedingly
rare.'' \citep[101]{rieder_scrutinizing_2017} To alleviate this issue,
\citet{rieder_scrutinizing_2017} suggests that ``algorithmic
technique'', which he refers to ``the finite set of well-known
approaches to information filtering and classification that underpin
most running systems'', can serve as a middle layer between high label
theorization and concrete implemented algorithmic systems. The former is
usually abstract and too macro. For example, in recent research on
digital capitalism, ``the particularities of these (algorithmic)
technological feats and how they relate to the political economy of `big
tech' remain underexplored.'' The latter are usually adaptive and
constantly changing under agile development, yielding auditing them from
outside mostly ineffective \citep{rieder_political_2022}.

The advantage of ``algorithmic technique'' is that they are close to
paradigm algorithms in the text-books, and it provides an accessible
language to describe the empirics, without ``turning to a singular
logic.'' This requires a researcher to get familiar with the technical
details of these techniques, partially through reading the computer
science literature as they are intended to be read, and partially
through hands-on these algorithms. Using the Bayes classifier as a
showcase, Rieder shows how this layer serves as a model to look at and
act in the world. Later, this approach was extended to analyze
``Transformer'', the building block of current large language models
\citep{luitse_great_2021}.

It is important to clarify that neither Mackenzie nor Rieder suggest
that analyzing operational formations or algorithmic techniques
eliminates the need for empirical research on actual algorithmic
systems. These two similar approaches delineate a space of possibilities
and impossibilities that constitute the actual algorithmic system in the
wild. In this sense, they can be seen as attempts to build ideal
types---conceptual models that provide researchers with a mediating
language for engaging with technical details in their empirics. These
ideal types also serve as reference points for comparing and contrasting
real-world systems with what are diagrammatically possible. By employing
them, researchers can better understand how algorithmic systems operate
within specific contexts.

In short, producing a thick description of algorithmic systems in
critical research requires engaging with their technical features. A
possible approach is to tinker with similar technologies, gaining
hands-on experience and exploring the subject positions behind the
constructing and maintenance of these systems.

\subsection{Offering new perspectives}\label{offering-new-perspectives}

Not only does tinkering prove to be valuable to approach algorithmic
systems, it also offers new perspectives on empirics of our broader
digital lives.

To further illustrate this point, we use anthropology---traditionally
considered a qualitative discipline---as an example in this section. In
recent years, computational methods also made its way into anthropology.
In the early stage of it, many anthropologists just mimic what other
computational social scientists are doing, and see machine learning
techniques like topic models as valuable tools to analyze the data. They
believe these methods based on codes/rules would make the interpretation
process more transparent and replicable.

\citet{munk_thick_2022} reject such a dichotomy of formalism and
interpretation assumed in the naive importing of computational methods.
Since modern machine learning is drastically different from rule-based
AI, they contend that applying computational methods is far more than
reintroducing formalism into the anthropology. Using a single predictive
model of emoji trained on a corpus of Facebook pages, they devised three
different games on it. The first game is a ``naive ethnscientist'' one,
of which the goal is to predict the users' usage of emoji as accurate as
possible. This game can be accessed with the same assessment paradigm in
predictive modeling. In the second ``reflexive ethnoscientist'' game,
the performances of the model is not compared with grounded truth, but
with ethnographers who try to imitate user of Facebook. This reveals the
differences between a machine learning model and trained ethnographers
in similar situations. Lastly, the ``interpretative ethnographer game''
focuses on where the machine learning algorithms fail to predict.
\citet{munk_thick_2022} suggest that machine learning is good at
capturing unambiguous cases. Thus, where machine learning produc the
wrong answer is more worth of thick description. In other words, the
ethnography begins where machine learning fails.

\citet{munk_thick_2022} further suggests that this approach fits the
post-hoc interpretability of machine learning. We read this article more
radical than this original intention, and see the three games devised
represent different configurations between the model and the world.
Given that the machine learning models are trained to make better
predictions, it has no inherent relation with the world
\citep{breiman_statistical_2001}. This is why the formalist
understanding (Or rule-based one) is not suitable with the machine
learning trained models in the first place: even if a model predicts
well, researchers can hardly make explication on the model directly.
Tinkering implies something else: diverse assessments of the model can
situate it uniquely in the world. We are not forcing the model for
better prediction, but trying to find a possible way to approach it.

This more radical approach resonates with what \citet{munk_thick_2022}
would still see as interpretive anthropology. In her work about energy
data technologies, \citet{knox_hacking_2021} places ``hacking'' at the
crux of her fieldwork. Here, hacking here is similar to tinkering data
technologies that the original scripts do not designate. Similarly,
\citet{jansky_digitized_2024} described similar collective tinkering of
data technologies in the Diabetes patient community.

Beyond that, \citet{knox_hacking_2021} contends that anthropologists
should participate in hacking themselves. She organized a series of
hackathons. These events not only created sites for further observation,
they also constitute public interventions into how technologies could
enact the world. More importantly for our purpose in this paper. Knox
suggests that ``hacking'' sheds light on a new mode of anthropology in
studying the data world. For \citet{knox_hacking_2021}, one thing stands
out : Hacking provides a way to speak back to the scalings of digital
data infrastructure via assembling possible stakeholders and
understanding how the world ``are made, formed, reinforced, and
dismantled,'' not via creating another scalable model.

In both cases, researchers need to get closer to the data practices, and
this affinity also provokes them to examine the empirics through new
lenses or configurations. This also brings us to the last point we would
like to make about tinkering as a way of knowing: it offers us a way to
capture the new cultural generalities.

\subsection{Tinkering as a new way of
intervening?}\label{tinkering-as-a-new-way-of-intervening}

Our discussion of tinkering culminates in what we see as its most urgent
significance: it offers a path toward an interventionist orientation in
algorithm studies. This is not merely a normative stance, but a
necessary response to our historical condition. Especially in the wake
of commercial and open-access large language models (LLMs), which have
dramatically reconfigured knowledge infrastructures within just a year
or two, we can no longer imagine a world without algorithms. These
systems are no longer external objects to be critiqued from a
distance---they constitute the very milieu in which we think, sense, and
live. As Alexander Galloway argues in Uncomputable, the question is no
longer how to model the world computationally, but how to inhabit a
world where computation itself sets the terms of inhabitation---one that
began much earlier than what we tend to mark as the beginning of the
digital age\citep{galloway_uncomputable_2021}.

The current wave of AI has triggered numerous debates about the boundary
between human judgment and computational intelligence. Yet as scholars
like Galloway, Mackenzie, and Seaver remind us, algorithmic systems are
not simply epistemic tools but affective and sensory infrastructures.
They shape how things are felt, anticipated, and acted upon. From this
perspective, drawing a boundary between human and machine is not an act
of clarity but of distortion. The challenge is not to determine where
``we'' end and ``the machine'' begins, but to understand how algorithmic
logic has already become constitutive of thought itself. As Heim already
observed in 1993, our habits of reading and inquiry---shaped by the
Boolean structures of search, consultation, and access---already reflect
an ongoing synbiosis between cognitive practice and computational
architecture \citep{heim1993metaphysics}. Thirty years onwards, can we
ever imagine again how we read only through books and letters without
Google search or Google scholar or PDF keyword search engines that are
just like the ink and pen in the old days? The question now is not
whether but how we have thought through machines.

This is why traditional distinctions between experts and non-experts
have begun to lose their meaning. Algorithms no longer belong to
specialized domains---they are entangled in ordinary experience,
generating effects that are ambient, relational, and distributed. They
produce object-worlds that are modular, dynamic, and
self-updating---resistant to totalization or external mastery. As Donna
Haraway has long urged, the only viable response to such conditions is
to stay with the trouble: to engage, to interfere, to
tinker\citep{haraway2016staying}. Tinkering, in this sense, is not an
attempt to fix from outside, but to generate proximity, improvisation,
and alternative relations from within.

Philosophically, this signals a shift from practical philosophy to a
realm of operational philosophy---one grounded not in a priori critique
or human practices alone, but in the entanglement of human life with
systems that generate their own conditions of being. As
\citet{mol_care_2010} has argued, tinkering is a modality of attentive
experimentation---a way of inhabiting the tension between critique and
care. Judgments of a technological system cannot be passed from outside;
failures to understand are often failures to dwell long enough in the
entangled folds of algorithmic life. In this sense, tinkering becomes a
way of intervening without mastery, of leaning into frictions and
failures to learn how to live differently with machines---and in doing
so, to learn something about ourselves as already embedded in these
worlds.

We see this approach as resonant with a longer anthropological tradition
of making as thinking. Here we are not referring merely to a narrow
disciplinary lineage, but to a broader human-making tradition. From
Lévi-Strauss's figure of the bricoleur to Tim Ingold's attention to the
textures of making, this lineage affirms the epistemic force of crafting
in and with the world
\citep{levi-strauss_savage_1966, ingold_footprints_2010}. Importantly,
these scholars did not invent this modality---they just gave name and/or
an intellectual form to a critical way of knowing where we are that have
long existed outside the domain of academic writing and Bacanian
empiricism. It is a modality that has quietly defined human existence:
too grounded, too tactile, too situated to sit comfortably within
dominant intellectualist paradigms. For those whose lives are not
organized around textual production, making is not metaphorical---it is
existential. If computation is imperial---seeking to tame chance, to
neutralize uncertainty, and to colonize the unknown---then tinkering, as
what makes possible a science of the concrete in Lévi-Straussian terms,
offers a grounded counter-politics. Human-centered thinking is, at its
core, making-centered thinking. In this sense, we are not opposing
algorithms per se, but insisting that the critical labor of inhabiting
uncertainty must not be outsourced---to computation or to abstraction.
We must feel our way through, with and against the systems that shape
our conditions of life. That is: we tinker, and in doing so, leave
traces---like footprints in a rain-soaked forest, as Ingold might say.
In a world where algorithmic systems are continuously shaping the
contours of our lifeworld, only a form of research that intervenes---and
intervenes with care---is itself the making of space for alternatives.

\section{Discussion and Conclusion: Tinkering as ways of making, knowing
and
caring}\label{discussion-and-conclusion-tinkering-as-ways-of-making-knowing-and-caring}

We propose that tinkering is not merely a methodological device but a
composite orientation --- simultaneously a mode of making, a way of
knowing and an act of caring.

As demonstrated in the examples above, tinkering offers a promising path
forward for both computational social science and critical algorithm
studies. We further argue that tinkering provides a point of
interventionist convergence where these two subfields can mutually
inspire and reshape one another. As we have shown, tinkering enables
critical researchers to engage deeply with existing algorithmic systems,
while much of the tooling and practical expertise needed for such
engagement currently resides in computational social science---where
they are not yet part of mainstream curricula. . The exchange, however,
must be mutual. as \citet{li_old_2025} and
\citet{ji_metasemanticmetapragmatic_2025} illustrate, critical
epistemologies can also reshape the practices of modeling itself.As
discussed elsewhere, this feedback loop opens up the possibility for
computational social science to be transformed from within
\citep{zhang_sts_2025}.

Tinkering also underscores the interpretive role of social sciences in
engaging with machine learning models--- beyond the narrow logics of
mechanical interpretability. Since predictive models are not meant to
mirror natural data-generating processes, their parameters, no matter
how ``simple'' the model ---lack substantive causal meaning. Both
\citet{li_old_2025} and \citet{munk_thick_2022} exemplify post-hoc
interpretations of models, while
\citet{ji_metasemanticmetapragmatic_2025} provides a new benchmark for
evaluating generative models. Collectively, these works treat models not
as purely technical artifacts but as cultural forms---open to
negotiation, situated within broader epistemic and normative landscapes.
.

Such an orientation demands a shift in how we relate to models.
Traditionally, researchers have approached modeling as a form of
extraction---either mining models for statistical significance or
milking them for higher accuracy. Tinkering challenges this coercive
view. Instead of imposing rigid expectations on models, tinkering
invites a more negotiated and responsive relation. It offers a form of
engagement that brings forth possibilities, rather than imposing rigid
outcomes. In this way, tinkering becomes a mode of dialogical alignment:
between researchers, models, and the social worlds in which they
operate.

Finally, we position, tinkering is part of a broader historical
transition in our relationships with technological systems..As a form of
care, tinkering resists the dominant logics of customization or
ordinalization that have dominated human-technology interactions over
the past two decades
\citep{lury_algorithmic_2019, fourcade_ordinal_2024}. It invites both
individuals and collectives to inhabit technology differently ---to work
within constraints, across frictions, and toward partial, situated forms
of repair. It asks not how we scale up, but how we stay close. And in
doing so, it gestures toward a social science---and a politics---that is
less about prediction and more about preservation, transformation, and
the everyday labor of making better worlds.

\bibliographystyle{apalike}
\renewcommand\refname{Bibliography}
\bibliography{references.bib}

\end{document}